# Synonymous Codon Usage Bias Overrides Phylogeny to Reflect Convergent Frond Architecture in a Rapidly Radiating Fern Family Thelypteridaceae

Kerui Huang, Wenyan Zhao, Huan Li, Ningyun Zhang, Lixuan Xiang, Xuan Tang, Yulong Xiao, Yi Liu, Zui Yao, Jun Yan, Hanbin Yin, Rongjie Huang, Yulong Xiao, Peng Xie, Haoliang Hu*, Jiangping Shu*, Hui Shang*, Yun Wang*

**Abstract**

Convergent evolution provides powerful evidence for natural selection, yet its molecular basis is typically sought in protein-coding amino acid substitutions. Whether adaptive pressures can drive the convergent evolution of synonymous codon usage bias (CUB) to override phylogenetic history remains a fundamental question. Here, we investigate this within the rapidly radiating fern family Thelypteridaceae by establishing a comparative framework that integrates chloroplast phylogenomics with dimensionality reduction of codon usage, morphological data, and divergence time estimation. Our results reveal that chloroplast CUB patterns are strikingly incongruent with the phylogeny of this family. Instead, they partition species into distinct clusters that strongly correlate with a convergently evolved morphological trait—lamina base architecture—a key adaptation whose radiation we date to the early Neogene. This convergent molecular signal is driven by a specific subset of photosynthesis-related genes (ndhJ, psaA, and psbD), which exhibit a high density of type-specific, third-position codon substitutions. These findings demonstrate that CUB can serve as a powerful, quantifiable indicator of adaptive history, revealing a cryptic layer of molecular convergence linked to the regulation of protein synthesis. Our work providing a new framework for uncovering adaptive histories obscured by complex evolutionary processes.

**Introduction**

Convergent evolution, the independent emergence of similar traits in different lineages, provides compelling evidence for the role of natural selection in shaping organismal form and function. Classic examples, from the streamlined bodies of sharks and dolphins to the complex sensory systems of echolocating bats and cetaceans, have long illustrated this principle at the morphological level[1]. In the genomic era, research has increasingly focused on uncovering the molecular underpinnings of these convergent phenotypes, often revealing parallel amino acid substitutions in key genes that directly alter protein function. Landmark studies have demonstrated this phenomenon in genes associated with toxin resistance across diverse animal phyla[2] and in the sensory pathways of marine mammals[3]. These investigations have primarily centered on non-synonymous mutations, as their impact on protein structure and function provides a direct link to adaptation. However, this focus means that a more subtle layer of molecular evolution, the evolution of synonymous sites, has received less attention.

Synonymous codon usage bias (CUB), the non-random use of codons for the same amino acid, is a pervasive genomic feature shaped by an interplay between neutral evolutionary forces, such as mutation pressure and genetic drift, and natural selection for translational optimization[4,5], .Because CUB is influenced by both ancestry-driven mutational biases and lineage-specific selective pressures, the prevailing expectation is that its patterns should largely mirror the organismal phylogeny. Yet, given that selection for translational efficiency and accuracy can be a potent evolutionary force, CUB may also reflect adaptive pressures that transcend phylogenetic history.

This raises a fundamental question: can strong, phenotype-driven selective pressures drive the convergent evolution of CUB to such an extent that it creates a genomic signature of adaptation that overrides the signal of ancestry? Answering this question would provide insight into a different mechanism of molecular convergence, linking adaptive evolution not to the protein product itself, but to the efficiency and regulation of its synthesis.

The fern family Thelypteridaceae presents an effective model system for investigating this question. As one of the largest and most cosmopolitan fern families, it has undergone a recent and rapid radiation, resulting in considerable morphological diversity, particularly in frond architecture[5,6]. Frond morphology in this family is not merely a taxonomic character but is known to be linked to ecological adaptation, with features like leaf dissection correlating with climatic variables[7],. For this purpose, the chloroplast genome provides an ideal molecular dataset. Its conserved gene content, lack of recombination, and uniparental inheritance offer a clear and tractable source of phylogenetic information against which adaptive signals can be compared[8,9], Crucially, as its genes are almost exclusively involved in photosynthesis and related metabolic processes, the chloroplast genome is functionally linked to the very adaptations—such as frond architecture for light capture[7]—that are central to our hypothesis. This makes it a powerful, self-contained system for investigating how selective pressures on organismal phenotype might be reflected in the evolution of codon usage. Furthermore, the family's evolutionary history is known to be complex, marked by unresolved generic relationships and a high incidence of morphological homoplasy[10,11], This combination of rapid diversification, morphological homoplasy, and complex evolutionary history makes Thelypteridaceae an ideal system for differentiating the signals of vertical descent from those of convergent adaptation.

In this study, we test the hypothesis that patterns of chloroplast codon usage bias in Thelypteridaceae are incongruent with the family's phylogeny and instead correlate with convergently evolved frond morphologies. To address this, we employed a multi-step analytical framework. First, we sequenced the complete chloroplast genomes of three key *Christella* species and constructed a robust, family-wide phylogeny based on 31 protein-coding genes, and estimated the divergence time based on this phylogeny. Second, we used Uniform Manifold Approximation and Projection (UMAP), a dimensionality reduction technique, to visualize the patterns of CUB and test for incongruence with the established phylogeny. Third, we correlated the resulting CUB-based clusters with morphological trait of lamina using the comprehensive taxonomic classification of Fawcett and Smith (2021)[12]and established a temporal framework for an important pattern through divergence time estimation. Finally, we performed a gene-by-gene analysis to identify the specific chloroplast genes driving the convergent molecular signal. Our findings reveal a notable instance of molecular convergence at the level of synonymous codons, directly linked to a major morphological adaptation, and identify a specific subset of photosynthesis-related genes as the primary drivers of this evolutionary pattern. This work provides compelling evidence that CUB can serve as a quantifiable indicator of adaptive history, capable of retaining a clear record of morphological convergence that is not apparent from phylogeny alone.

## Methods
**Plant materials, chloroplast genome sequencing, assembling and annotation**

Fresh leaves were picked from *Christella acuminatus*, *Christella parasiticus*, and *Christella latipinnus* (Figure 1), from plants naturally growing in Hengshan, Hengyang, Hunan province, China (N27°16'58.523", E112°42'34.654", 702 m). The voucher specimens were deposited at the College of Life and Environmental Sciences, Hunan University of Arts and Sciences (Contact Person: Kerui Huang, huangkerui008@163.com, voucher numbers HS005, HS006, HS007, HS008).

The library was constructed using the DNAsecure Plant Kit (TIANGEN Biotech Co., Ltd., Beijing) and the sequencing was performed on an Illumina HiSeq 2500 platform (San Diego, CA), both outsourced to Shanghai Personalbio Technology Co., Ltd. (China)

After filtering out the low-quality reads using fastp[13] clean reads were retained for further analysis. The chloroplast genome of the three species were de novo assembled using GetOrganelle v1.7.5[14] with parameters set as -R 15 -k 21,45,65,85,105 -F embplant_pt. Subsequently, the assembled chloroplast genome was annotated using CPGAVAS2[15] with default settings, and a circular genome map was visualized using CPGView (http://www.1kmpg.cn/cpgview/).

**Phylogenetic analysis**

The GenBank files for a total of 37 chloroplast genomes, comprising all available Thelypteridaceae species from NCBI (including three newly characterized in this study) and two Woodsia species as outgroups, were downloaded for phylogenetic analysis. Among them, 31 protein-coding genes shared by all genomes were screened out for subsequent analysis. Sequence alignment of each gene was performed separately using MAFFT v7.313[16]. Gblocks 0.91b was then utilized to remove poorly aligned regions of each gene. The filtered gene sequences were concatenated head-to-tail into supergenes[17]. Maximum likelihood phylogenies were generated using IQ-TREE v1.6.12[18]. The GTR+F+I+G4 model was selected based on the Bayesian Information Criterion (BIC) in ModelFinder. This process was further strengthened with 5000 ultrafast bootstrap replications for robust statistical support along with Shimodaira-Hasegawa-like approximate likelihood ratio test.

**Divergence time estimation**

The divergence times for the species included in our phylogenetic analysis were estimated using the Markov chain Monte Carlo (MCMC) approach implemented in the PAML software package, specifically utilizing its MCMCtree program[19]. The optimal phylogenetic tree topology for our dataset was determined using IQ-TREE.

For calibrating the molecular clock, we incorporated three fossil-based calibration points derived from previous studies. These calibration points were as follows: (F1) between 48.6 and 103.7 million years ago (Ma)[20-25], (F2) between 44.6 and 53.0 Ma[24,25], and (F3) between 20.0 and 27.7 Ma[23,24]. These points were strategically chosen to constrain each corresponding node in the phylogenetic tree.

Our analysis employed the independent rates model (IRM), which assumes a lognormal distribution for rate variation among lineages. The JTT model was selected for amino acid substitution, with the alpha parameter for gamma-distributed rate variation across sites set at 0.5, using 5 discrete categories. The analysis was performed using the approximate likelihood method. The birth-death process model was used to establish priors for node ages, with the birth rate ($\lambda$) and death rate ($\mu$) both set to 1, and the sampling proportion (s) set to 0.1. For the

MCMC analysis, the initial 10% of trees generated were discarded as burn-in. Subsequent trees were sampled every 10 iterations, culminating in a total of 10,000 sampled trees for the final analysis.

**Codon Usage Bias Analysis**

Annotated protein-coding genes were extracted from each plastome and curated to ensure complete reading frames (length divisible by three, no internal stop codons). Only loci present in every species entered the comparative dataset; any accession lacking a gene removed that locus from the shared set. For each retained gene, sequences were exported in FASTA format with identifiers concatenating species and gene names to preserve provenance while remaining compatible with subsequent alignment and parsing steps.

For every gene in every accession we calculated an extensive panel of codon bias metrics: positional GC content (GC1/2/3), nucleotide frequencies at synonymous third positions (A3, T3, C3, G3), global GC and GC3s, the effective number of codons (ENC), codon adaptation index (CAI), frequency of optimal codons (Fop), parity rule 2 (PR2) coordinates, neutrality plot components, and relative synonymous codon usage (RSCU) for all 59 sense codons. Standard formulas from codon usage theory[26] were implemented in Python 3.11 with Biopython 1.81 and CodonW-compatible routines, ensuring consistency with plastid genetic code expectations. Species-by-gene metric tables and accompanying RSCU matrices were written to comma-separated files for use in multivariate analyses.

**Codon Usage UMAP Visualization**

Feature matrices were z-score standardized (within each column) and flattened so that every species was represented by a single high-dimensional vector concatenating metrics across all shared genes. Uniform Manifold Approximation and Projection (UMAP; umap-learn 0.5) provided a nonlinear embedding of this space. A grid search explored n_neighbors values of 5, 15, 30, and 50, min_dist of 0.0, 0.1, 0.3, and 0.5, and Euclidean versus correlation distance metrics. Silhouette coefficients computed against the pre-defined lamina architecture classes identified the optimal hyperparameters, after which the best two-dimensional embedding was rendered interactively with Plotly to visualize correspondence between codon usage clusters and frond morphology. An analogous workflow applied to gene-wise RSCU matrices to identify genes with outsized influence on the separation.

**Type-Specific Synonymous Site Detection**

For each shared gene, multiple-sequence alignments were generated with MAFFT v7 (--auto strategy, full iterative refinement) and inspected to confirm co-linear alignments with uniform lengths that remained multiples of three, preventing codon frame shifts. Variable sites were enumerated, and a consensus-within-types algorithm classified a position as type-specific when every species inside a type shared the same nucleotide while this nucleotide differed between types. A relaxed mode allowed designated heterogeneous types to retain multiple nucleotides provided none overlapped the strict-Type Consensus, capturing biologically meaningful but noisy patterns. Counts of variable versus specific sites and per-type tallies were recorded for each gene.

## Results
### 1. Chloroplast Genome Assembly, Annotation, and Characteristics

High-throughput sequencing and assembly yielded the complete chloroplast (cp) genomes for the three species, which were *C. acuminatus* (Figure 1a), *C. parasiticus* (Figure 1b), and *C. latipinnus* (Figure 1c). The cp genomes of *C. acuminatus*, *C. parasiticus*, and *C. latipinnus* exhibited the typical quadripartite circular structure characteristic of most land plant plastomes, comprising a large single-copy (LSC) region, a small single-copy (SSC) region, and two inverted repeat (IRa and IRb) regions (Figure 1). The genomic features, including total size, regional lengths, gene content, and GC distribution, were highly conserved across the three species. Detailed statistics are presented in Table 1. Overall, these features are consistent with those reported for other Thelypteridaceae species, such as *Christella dentata*[27], and generally align with typical characteristics of fern and other land plant chloroplast genomes[8,9], The successful assembly and annotation provided a robust foundation for subsequent phylogenetic and comparative genomic analyses.

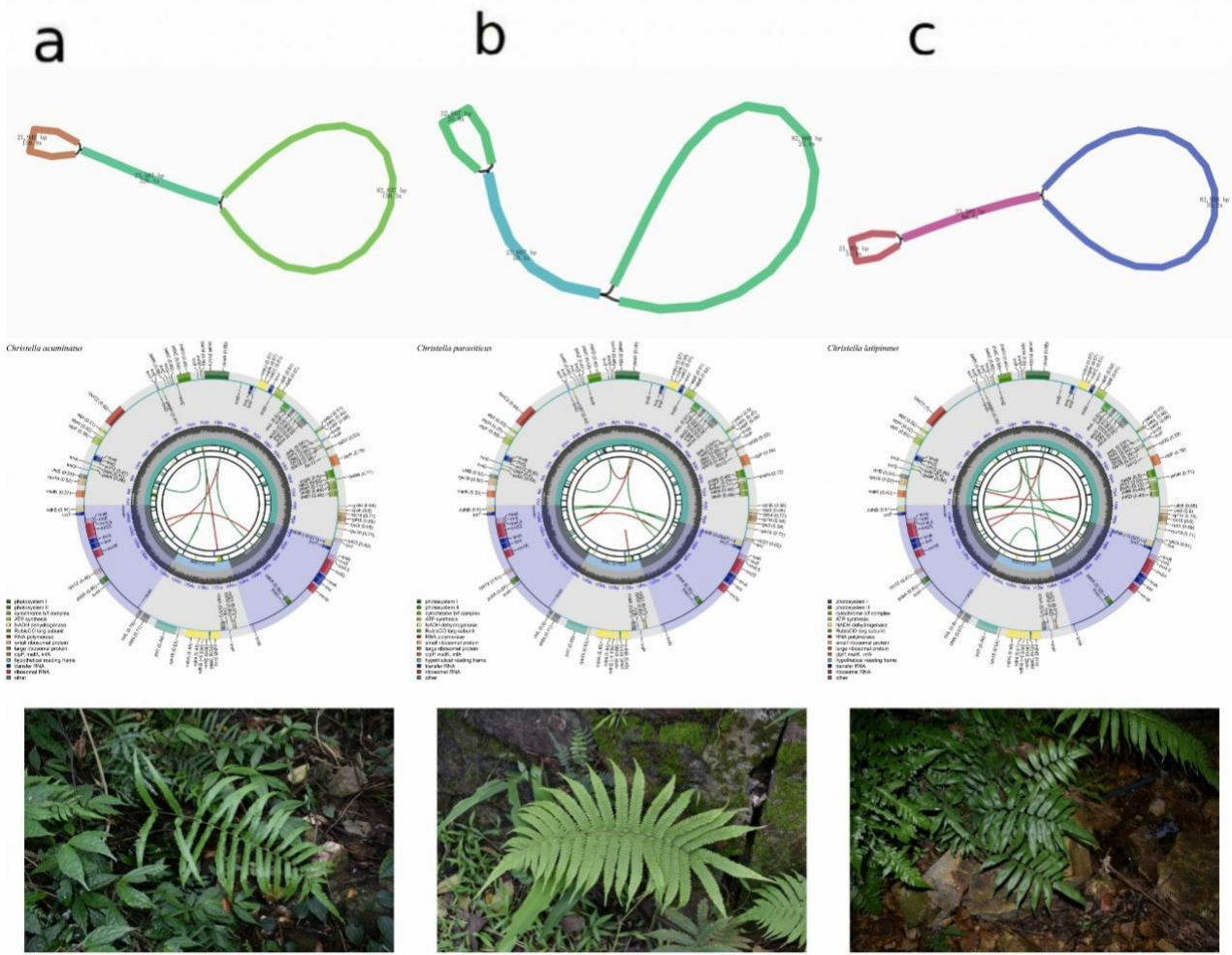

**Figure 1.** Chloroplast Genome Assembly, Annotation, and Morphology of Three Newly Sequenced *Christella* Species.

This figure provides a comprehensive overview of the chloroplast genomes for (a) *Christella acuminatus*, (b) *Christella parasiticus*, and (c) *Christella latipinnus*. The top row presents Bandage assembly graphs that confirm the complete circular structure of each plastome, with annotations indicating contig length and average sequencing coverage. The middle row displays detailed circular gene maps illustrating the typical quadripartite architecture composed of the Large Single-Copy (LSC), Small Single-Copy (SSC), and two Inverted Repeat (IRa and IRb) regions. Within these maps, genes are color-coded according to their functional groups, and their transcriptional direction is indicated by their position relative to the central ring. An inner track plots the variation in GC content across the genome. The bottom row provides photographs of the living plants.

**Table 1.** Chloroplast Genome Features of the Three Newly Sequenced Christella Species

| Species | Total Length (bp) | LSC Length (bp) | SSC Length (bp) | IR Length (bp) | Unique Genes | PCGs | tRNA Genes | rRNA Genes | Genes in IR | Overall GC (%) | IR GC (%) | LSC GC (%) | SSC GC (%) |
|---|---|---|---|---|---|---|---|---|---|---|---|---|---|
| Christella acuminatus | 151533 | 82627 | 21732 | 23587 | 127 | 82 | 37 | 8 | 13 | 42.44 | 45.07 | 41.75 | 39.32 |
| Christella parasiticus | 151573 | 82606 | 21753 | 23607 | 127 | 82 | 37 | 8 | 13 | 42.49 | 45.14 | 41.83 | 39.26 |
| Christella latipinnus | 151588 | 82694 | 21724 | 23585 | 127 | 82 | 37 | 8 | 13 | 42.43 | 45.13 | 41.75 | 39.15 |

**2. A Well-Supported Chloroplast Phylogeny for Thelypteridaceae**

To elucidate the evolutionary relationships of the species within Thelypteridaceae, a phylogenetic analysis was conducted using a concatenated dataset of 31 common protein-coding genes extracted from the chloroplast genomes of four samples representing three species sequenced in this study and from 31 additional Thelypteridaceae samples, which together constituted a total of 27 Thelypteridaceae species, representing all currently available Thelypteridaceae samples with chloroplast genomes, with two Woodsiaceae species as outgroups (Table 2).

The Maximum Likelihood (ML) phylogeny constructed using IQ-TREE under the GTR+F+I+G4 model, revealed a well-supported tree (Figure 2). The majority of nodes received bootstrap support over 90%, with many at 100% and all nodes supported by values over 50%, indicating the reliability of the inferred relationships. The overall topology of the phylogenetic tree was consistent with previous comprehensive classifications and molecular studies of Thelypteridaceae[6,28]. The family was clearly divided into two major monophyletic clades, corresponding to the subfamilies Phegopteridoideae (comprising Macrothelypteris, Phegopteris, and Pseudophegopteris) and Thelypteridoideae. Within Thelypteridoideae, several tribes and subtribes were resolved. Notably, *Cyclogramma* and *Stegnogramma* clustered together, forming the Leptogrammeae tribe, consistent with findings from previous studies[6,29]. The Meniscieae tribe included the subtribes Pseudocyclosorinae and Menisciinae. The Menisciinae subtribe was represented by a clade containing *Cyclosorus* and *Ampelopteris*. The Pseudocyclosorinae subtribe formed a large clade encompassing *Christella*, *Abacopteris*, *Pseudocyclosorus*, *Mesopteris*, *Grypothrix*, *Menisciopsis*, and *Glaphyropteridopsis*. Crucially, the genus *Christella*, including the three newly sequenced species (*C. acuminatus*, *C. parasiticus*, and *C. latipinnus*), was recovered as a strongly supported monophyletic group (BS=100%), supporting the congeneric status of these species and the traditional morphological classification (Figure 2). However, a notable incongruence was observed within *Christella parasiticus*: our newly sequenced sample (NC070301) did not cluster with a previously sequenced accession of the same species (MT130643). Repeated verification confirmed the correct identification of our specimen, indicating that this discrepancy reflects genuine biological complexity rather than data error. This finding is consistent with taxonomic revisions that describe *C. parasiticus* as a widespread and highly variable species complex with multiple ploidy levels and a history of hybridization[30], and aligns with recent phylogenomic studies that have demonstrated the prevalence of deep, intergeneric hybridization within the christelloid clade[10]. As the focus of this study is on adaptive evolution at the generic level, this intraspecific complexity does not affect our main conclusions but warrants further investigation.

**Table 2.** Thelypteridaceae species and related information used for constructing phylogenetic trees

| Species name | NCBI |
| --- | --- |
| *Ampelopteris prolifera* | NC035835 |
| *Ampelopteris prolifera* | MT130611 |
| ***Christella acuminatus*** | NC070299 |
| ***Christella acuminatus*** | NC070302 |
| *Christella dentatus* | PP236761 |
| *Christella dentatus* | NC061697 |
| ***Christella latipinnus*** | NC070300 |
| *Christella parasiticus* | NC070301 |
| *Christella procerus* | NC035842 |
| *Christella procurrens* | MT130695 |
| *Christella truncatus* | MT130565 |
| ***Christella parasiticus*** | MT130643 |
| *Cyclogramma auriculata* | MT130552 |
| *Cyclosorus interruptus* | NC057240 |
| *Glaphyropteridopsis erubescens* | MT130562 |
| *Glaphyropteridopsis erubescens* | MM623355 |
| *Grypothrix triphylla* | MN623361 |
| *Macrothelypteris torresiana* | MH500230 |
| *Macrothelypteris torresiana* | MT130591 |
| *Macrothelypteris torresiana* | NC035858 |
| *Mesopteris tonkinensis* | NC041428 |
| *Amauropelta beddomei* | MT130660 |
| *Coryphopteris japonica* | MT130553 |
| *Phegopteris decursive-pinnata* | MT130548 |
| *Phegopteris decursive-pinnata* | MN623353 |
| *Abacopteris gymnopteridifrons* | MT130561 |
| *Abacopteris gymnopteridifrons** | MT130555 |
| *Abacopteris penangianum* | MT130694 |
| *Trigonospora ciliata* | MT130659 |
| *Pseudocyclosorus esquirolii* | MT130607 |
| *Pseudophegopteris aurita* | NC035861 |
| *Pseudophegopteris pyrrhorhachis* | MT130575 |
| *Pseudophegopteris yunkweiensis* | MT130680 |
| *Stegnogramma griffithii* | MT130604 |
| *Stegnogramma sagittifolia* | NC035863 |
| *Woodsia macrochlaena* | NC035864 |
| *Woodsia polystichoides* | NC035865 |

Note: Species newly sequenced and characterized in this study are indicated in bold. * This accession is deposited in NCBI as *Grypothrix megacuspe*, but our phylogenetic analysis suggests it is likely misidentified and should be classified as *Abacopteris gymnopteridifrons*.

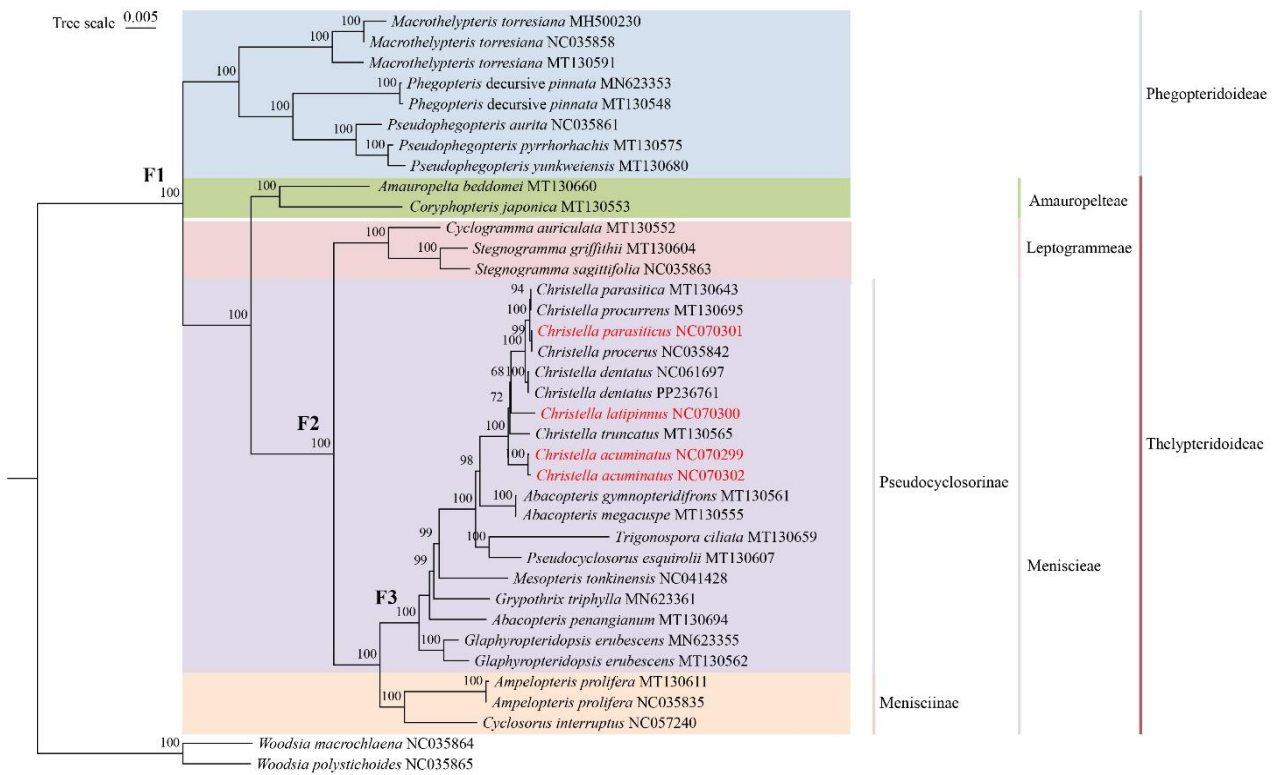

**Figure 2.** Phylogenetic Relationships within Thelypteridaceae Based on 31 Chloroplast Protein-Coding Genes. The Maximum Likelihood (ML) phylogeny was inferred from a concatenated alignment of 31 shared protein-coding genes. Bootstrap support values are shown at the nodes. Two species from Woodsiaceae were used as outgroups. Calibration points F1, F2, and F3, used for divergence time analysis, are marked on the tree with bold dots.

## 3. UMAP Analysis Reveals Codon Usage Patterns Are Strongly Incongruent with Phylogeny

Analysis of the coding sequences from the 37 Thelypteridaceae chloroplast genomes revealed distinct patterns of codon usage bias (CUB). The overall GC content of the coding sequences (GCall) across the analyzed species was consistently below 50%, ranging from 43.20% to 45.30% (Table 3). A characteristic trend in GC content at the three codon positions was observed: the first position (GC1) had the highest GC content, followed by the second position (GC2), and the third position (GC3) consistently exhibited the lowest GC content (GC1 > GC2 > GC3). For example, in *C. acuminatus* (NC070299), GC1 was 48.81%, GC2 was 42.12%, and GC3 was 39.17%. This pattern, particularly the low GC3 content, indicates a strong preference for A/T-ending codons, a feature commonly observed in plant chloroplast genomes 21[6,31-34]. The Effective Number of Codons (ENc) values for the Thelypteridaceae species ranged from 56.06 (*C. parasiticus*) to 57.17 (*Amauropelta beddomei*), suggesting a generally weak to moderate overall CUB (Table 3).

**Table 3.** Codon Usage Bias Characteristics of Chloroplast Genomes Coding Sequences in Thelypteridaceae

| Species | GenBank Accession | GCall | GC1 | GC2 | GC3 | GC3s | ENC |
|---|---|---|---|---|---|---|---|
| *Ampelopteris prolifera* | MT130611 | 43.70% | 49.17% | 42.46% | 39.30% | 36.80% | 56.24 |
| *Ampelopteris prolifera* | NC035835 | 44.60% | 50.11% | 43.48% | 40.04% | 37.40% | 56.38 |
| *Christella acuminatus* | NC070299 | 43.40% | 48.81% | 42.12% | 39.17% | 36.70% | 56.22 |
| *Christella acuminatus* | NC070302 | 43.40% | 48.80% | 42.11% | 39.17% | 36.70% | 56.22 |
| *Christella dentatus* | NC061697 | 43.40% | 48.45% | 42.45% | 39.15% | 36.70% | 56.19 |
| *Christella dentatus* | PP236761 | 44.70% | 50.95% | 42.30% | 40.54% | 37.80% | 56.40 |
| *Christella latipinnus* | NC070300 | 43.20% | 48.53% | 41.92% | 38.77% | 36.30% | 56.11 |
| *Christella parasiticus* | MT130643 | 43.20% | 48.58% | 41.90% | 38.80% | 36.30% | 56.07 |
| *Christella parasiticus* | NC070301 | 43.20% | 48.58% | 41.90% | 38.80% | 36.30% | 56.06 |
| *Christella procerus* | NC035842 | 44.60% | 49.96% | 43.45% | 40.25% | 37.60% | 56.52 |
| *Christella procurrens* | MT130695 | 43.20% | 48.58% | 41.90% | 38.80% | 36.30% | 56.07 |
| *Christella truncatus* | MT130565 | 43.60% | 48.90% | 42.27% | 39.28% | 36.80% | 56.25 |
| *Cyclogramma auriculata* | MT130552 | 43.90% | 49.20% | 42.56% | 39.67% | 37.20% | 56.57 |
| *Cyclosorus interruptus* | NC057240 | 43.60% | 48.38% | 42.66% | 39.66% | 37.00% | 56.14 |
| *Glaphyropteridopsis erubescens* | MN623355 | 44.90% | 50.19% | 43.68% | 40.54% | 37.90% | 56.68 |
| *Glaphyropteridopsis erubescens* | MT130562 | 44.00% | 49.42% | 42.50% | 39.86% | 37.40% | 56.49 |
| *Grypothrix triphylla* | MN623361 | 45.00% | 50.35% | 43.67% | 40.78% | 38.20% | 56.79 |
| *Menisciopsis penangianum* | MT130694 | 43.90% | 49.22% | 42.33% | 39.77% | 37.30% | 56.57 |
| *Macrothelypteris torresiana* | MH500230 | 45.10% | 50.55% | 43.17% | 41.26% | 38.70% | 56.89 |
| *Macrothelypteris torresiana* | MT130591 | 44.20% | 49.35% | 42.87% | 40.24% | 37.90% | 56.96 |

| Species | Accession | | | | | | |
|---|---|---|---|---|---|---|---|
| *Macrothelypteris torresiana* | NC035858 | 45.10% | 50.40% | 43.44% | 41.24% | 38.70% | 56.75 |
| *Mesopteris tonkinensis* | NC041428 | 44.90% | 50.28% | 43.58% | 40.63% | 38.00% | 56.70 |
| *Amauropelta beddomei* | MT130660 | 44.60% | 49.41% | 42.89% | 41.11% | 38.80% | 57.17 |
| *Coryphopteris japonica* | MT130553 | 44.20% | 48.98% | 42.44% | 40.84% | 38.40% | 57.10 |
| *Phegopteris decursive-pinnata* | MN623353 | 44.70% | 49.69% | 43.32% | 40.73% | 38.00% | 56.44 |
| *Phegopteris decursive-pinnata* | MT130548 | 43.80% | 49.10% | 42.20% | 39.85% | 37.30% | 56.50 |
| *Abacopteris gymnopteridifrons* | MT130561 | 43.80% | 49.25% | 42.13% | 39.65% | 37.10% | 56.46 |
| *Abacopteris gymnopteridifrons* | MT130555 | 43.80% | 49.25% | 42.13% | 39.65% | 37.10% | 56.47 |
| *Trigonospora ciliata* | MT130659 | 43.40% | 49.08% | 42.03% | 38.93% | 36.40% | 56.34 |
| *Pseudocyclosorus esquirolii* | MT130607 | 43.30% | 48.70% | 41.88% | 39.10% | 36.70% | 56.44 |
| *Pseudophegopteris aurita* | NC035861 | 45.10% | 50.29% | 43.38% | 41.36% | 38.80% | 56.74 |
| *Pseudophegopteris pyrrhorhachis* | MT130575 | 44.20% | 49.26% | 42.52% | 40.69% | 38.30% | 56.93 |
| *Pseudophegopteris yunkweiensis* | MT130680 | 44.30% | 49.51% | 42.52% | 40.72% | 38.30% | 56.92 |
| *Stegnogramma griffithii* | MT130604 | 44.10% | 49.22% | 42.40% | 40.51% | 38.10% | 56.70 |
| *Stegnogramma sagittifolia* | NC035863 | 45.30% | 50.67% | 43.71% | 41.33% | 38.70% | 56.77 |
| *Woodsia macrochlaena* | NC035864 | 45.10% | 50.65% | 43.18% | 41.12% | 38.40% | 56.97 |
| *Woodsia polystichoides* | NC035865 | 44.90% | 50.59% | 43.21% | 40.60% | 37.90% | 56.75 |

To visualize the complex landscape of CUB, Uniform Manifold Approximation and Projection (UMAP) was applied to a comprehensive set of codon usage parameters, where various codon preference indices for all protein-coding genes of each species (GC1, GC2, GC3, A3, T3, C3, G3, Nc, CAI, Fop, GC, GC3s, RSCU for each codon) were considered in this analysis. The global UMAP analysis, including species from Thelypteridaceae and Woodsia as an outgroup, revealed three distinct clusters (Type A, Type B, and Type C) (Figure 3). Type A comprised *Christella*, *Pseudocyclosorus*, and *Abacopteris*. Type B included *Pseudophegopteris*, *Phegopteris*, *Woodsia*, and *Macrothelypteris*. Type C was the largest and most diverse, containing *Mesopteris*, *Cyclosorus*, *Ampelopteris*, *Stegnogramma*, *Grypothrix*, *Abacopteris*, *Glaphyropteridopsis*, *Cyclogramma*, *Amauropelta* and *Coryphopteris*. Type A and Type C together constituted the large subfamily Thelypteridoideae.

To assess the extent to which these preference patterns reflect the evolutionary history of the species, we compared the codon preference patterns revealed by the global UMAP analysis (Figure 4b) with the phylogenetic tree constructed based on chloroplast genomes (Figure 4a). Overall, there was a certain degree of correspondence between the two. For example, Type B in UMAP roughly encompassed the clade located at relatively basal positions or representing earlier divergence events in the phylogenetic tree, the subfamily Phegopteridoideae (*Phegopteris*, *Pseudophegopteris*, *Macrothelypteris*), effectively distinguishing them from the core groups within Thelypteridoideae that underwent subsequent large-scale radiation (Type A and Type C in UMAP). Furthermore, *Pseudocyclosorus*, *Abacopteris*, and *Christella*, which clustered together in the phylogenetic tree, and all belong to the subtribe Pseudocyclosorinae were classified as Type A in UMAP, clearly separated from other types, reflecting consistency with phylogeny. However, significant discrepancies were mainly observed within the core groups of the Thelypteridoideae subfamily. The phylogenetic tree clearly divided this subfamily into well-supported monophyletic groups such as the Leptogrammeae tribe, Pseudocyclosorinae subtribe, and Menisciinae subtribe, but the UMAP clustering did not strictly follow this phylogenetic framework. Specifically, Type A in UMAP was only a subset of the Pseudocyclosorinae subtribe, while Type C in UMAP appeared as a "mixture," with its members widely sourced from phylogenetically distinct branches (Amauropelteae tribe, Leptogrammeae tribe, Pseudocyclosorinae subtribe (in part), and Menisciinae subtribe). This phenomenon, where codon preference clustering does not strictly adhere to phylogenetic boundaries, particularly the extensive mixing in Type C, is a notable departure from the expectation that CUB patterns largely mirror evolutionary history[31] and strongly suggests that phylogenetic relationships alone are insufficient to explain the observed codon preference patterns. Other evolutionary factors, such as convergent selection or specific genomic characteristics, may have played important roles in shaping the codon usage patterns of these core groups.

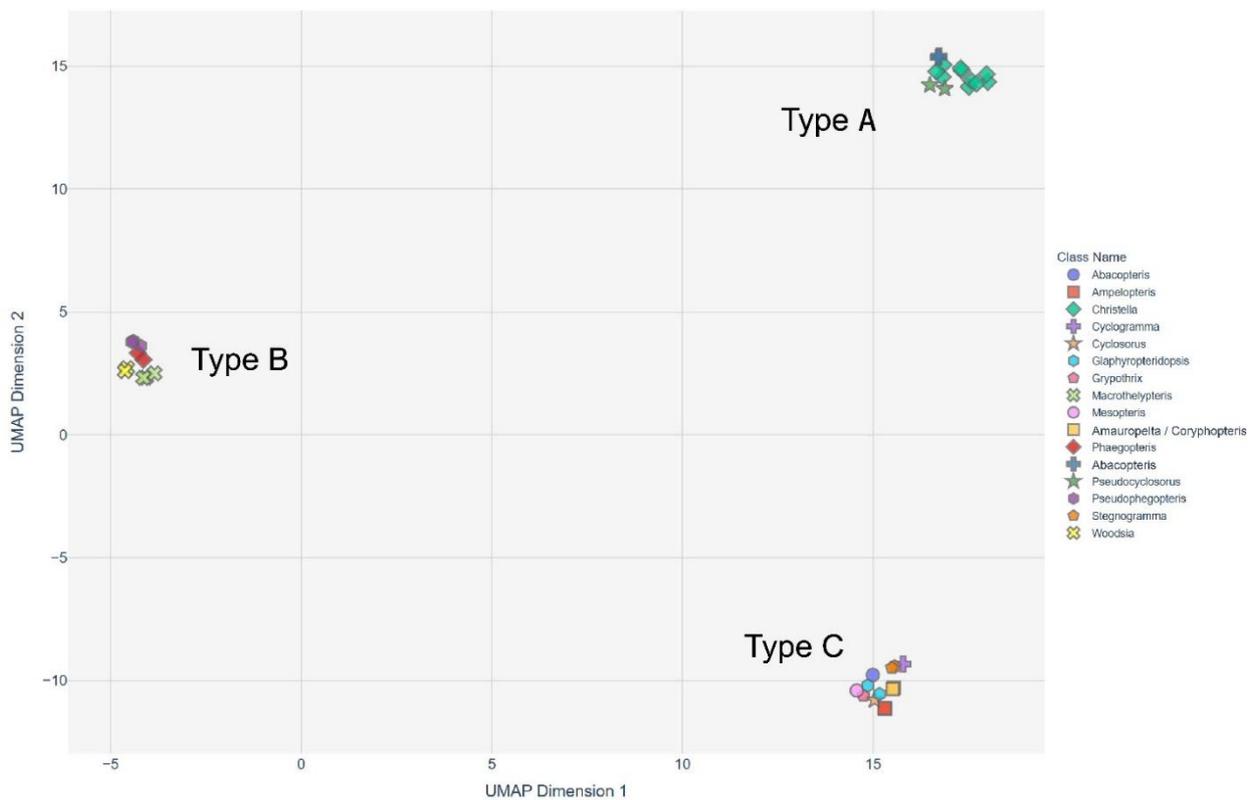

**Figure 3.** The Global UMAP Visualization of Codon Usage Bias among Thelypteridaceae species.

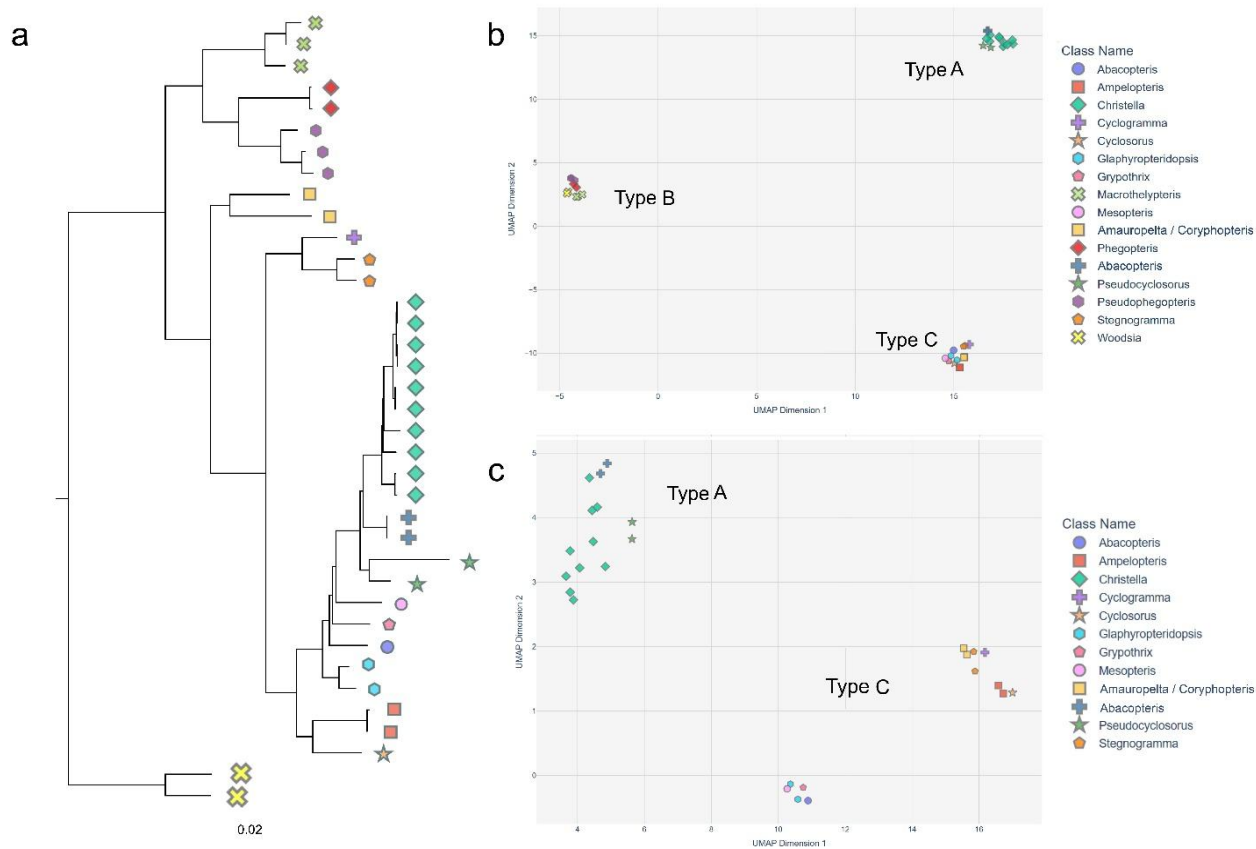

**Figure 4.** UMAP Visualization of Codon Usage Bias in Comparison with Phylogenetic Relationships. (a) Simplified Phylogenetic tree; (b) Global UMAP visualization; (c) Focused UMAP visualization.

Due to the observed inconsistencies between codon preference patterns in the global UMAP analysis and phylogenetic relationships, especially the conflicts evident in the core members of the Thelypteridoideae subfamily represented by Type A and Type C, we conducted a focused analysis to delve deeper into the subtle codon usage preference relationships within and among these types and attempt to resolve this inconsistency (Figure 4c). This analysis excluded all species from Type B and selected only genera belonging to the original Type A and Type C, using fewer species to obtain higher-resolution results based on more shared protein-coding genes. The results are presented in Figure 4c. It is evident that in the high-resolution results, the genera originally belonging to Type A and Type C remained separate and still clearly clustered into two major types. *Pseudocyclosorus*, *Abacopteris*, and *Christella* were still classified as Type A, while the Amauropelteae tribe, Leptogrammeae tribe, Pseudocyclosorinae subtribe (in part), and Menisciinae subtribe comprised Type C. This reinforces the findings of the global UMAP analysis, indicating that the separation between Type A and Type C is a robust and consistent feature of the dataset.

## 4. CUB-Based Clusters Correlate with a Convergently Evolved Lamina Architecture Originating in the Early Neogene

Our preceding analysis revealed a significant incongruence between the UMAP clustering based on codon usage bias and the phylogenetic relationships of the species of Thelypteridoideae subfamily represented by Type A and Type C. This strongly suggests that evolutionary driving forces beyond phylogeny, such as convergent evolution, may have played a significant role in shaping these genomic features.

To investigate the biological significance of the CUB-based clusters, we conducted a detailed morphological review of the taxa in Type A and Type C based on the comprehensive classification system of Fawcett and Smith (2021)[12]. This analysis revealed a strong correlation between the UMAP clusters and a key, convergently evolved morphological trait: lamina base architecture. The results showed that Type A, comprising *Abacopteris*, *Pseudocyclosorus*, and *Christella*, is uniformly characterized by a non-truncate, tapering lamina base (Table 4). This morphology is a direct consequence of the significant reduction of the proximal pinnae, a key diagnostic feature for these genera as documented in classical taxonomic studies[30,35].

Conversely, the phylogenetically diverse Type C is unified by the presence of a truncate lamina base (Table 4). This type, which includes genera from different tribes such as *Cyclosorus*, *Glaphyropteridopsis*, *Stegnogramma*, *Mesopteris*, *Grypothrix*, *Cyclogramma*, *Ampelopteris*, *Abacopteris*, *Amauropelta* and *Coryphopteris*, is defined by having proximal pinnae that are not, or are only slightly, reduced. As a result, the laminae of these genera are broadest at or near the base. This morphological trait is consistently highlighted across these otherwise diverse taxa in authoritative taxonomic treatments[7,12,36]. Therefore, the high degree of correlation between the codon usage patterns revealed by UMAP clustering and the lamina base architecture provides a strong morphological correlate for the previously observed incongruence between CUB and phylogeny.

**Table 4.** Comparison of Lamina Base Morphology Between Type A and Type C of Thelypteridaceae

| Types | Genus | Lamina Base Morphology |
|---|---|---|
| Type A | *Abacopteris* | Non-truncate; Proximal pinnae are not or only slightly reduced. |
| | *Pseudocyclosorus* | Non-truncate; Tapering. Characterized by proximal pinnae that are abruptly reduced into a series of opposite, auriculate pinnae, or even to peg-like aerophores without lamina tissue. |
| | *Christella* | Non-truncate; Gradually Tapering. A key diagnostic feature is the presence of 1 to 10 pairs of proximal pinnae that are gradually reduced in size, creating an attenuate base. |
| Type C | *Cyclosorus* | Truncate. Proximal pinnae are not reduced, or the lowermost pair is only slightly so. |
| | *Glaphyropteridopsis* | Truncate. Described as having "truncate bases (lacking greatly reduced pinnae)." |
| | *Stegnogramma* | Truncate. Blades are often widest at or near the base, with proximal pinnae never being significantly reduced. |
| | *Mesopteris* | Truncate. Proximal pinnae are not or little reduced. |
| | *Grypothrix* | Truncate. Proximal pinnae are not reduced. |

| | |
|---|---|
| *Cyclogramma* | Truncate. Proximal pinnae are not noticeably or only slightly shorter than the medial ones. (A few species may show some reduction, but it is not the defining characteristic of the genus). |
| *Ampelopteris* | Truncate. Proximal pinnae of well-developed fronds are not reduced or only a few pairs are slightly reduced. |
| *Abacopteris* | Truncate. Proximal pinnae are not or only slightly reduced. |
| *Amauropelta* and *Coryphopteris* [1] | Truncate. Proximal pinnae are described as "abruptly- or little reduced". |

To place the emergence of this key morphological innovation within a temporal framework, we performed a divergence time estimation. The resulting chronogram (Figure 5) reveals that the radiation of the Type A clade, which is unified by this tapering lamina base morphology, began approximately 19.43 Ma. This timing places the diversification of this architectural syndrome in the early Neogene (Miocene), a period of significant global climate change. Therefore, the strong correlation between codon usage patterns and lamina base architecture is further supported by a distinct temporal origin, providing evidence that the CUB signal may reflect an adaptive radiation event rather than solely phylogenetic history.

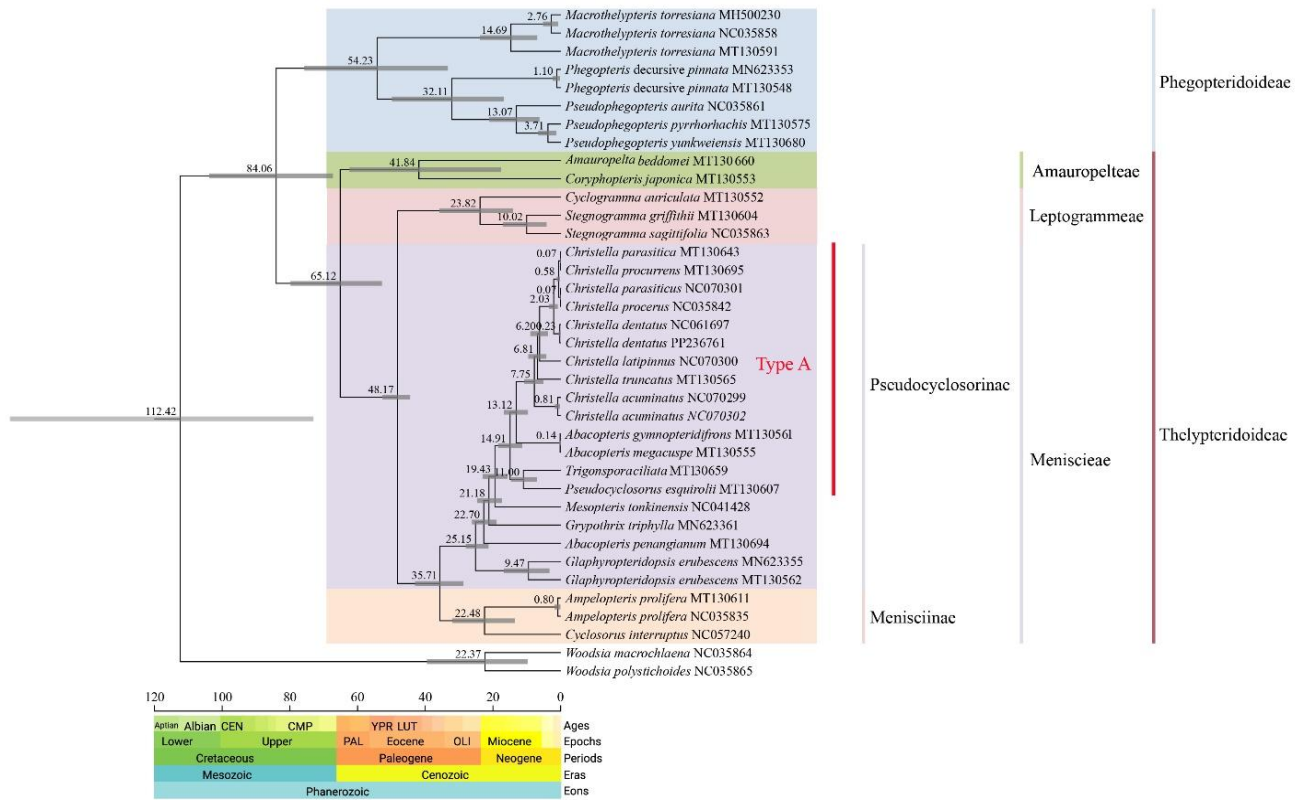

**Figure 5.** Divergence time estimation based on cp genome sequences. The divergence times are exhibited on each node, whereas the greed bars represent the 95% highest posterior density interval for each node age.

## 5. Natural Selection is the Predominant Force Shaping Codon Usage

To determine the evolutionary forces driving CUB in Thelypteridoideae, we performed ENC-plot, PR2-plot, and neutrality plot analyses on representative species from Type A and Type C. The ENC-plot results showed that while most genes were distributed near the standard curve, a few deviated, suggesting that CUB is influenced by factors beyond simple mutational bias (Figure 6a). Notably, the *psbA* gene in every species was located distinctly below the standard curve, indicating that this gene exhibits a particularly strong codon preference in both Type A and Type C species. The PR2-plot analysis revealed a non-uniform distribution of points, primarily in the G3/(G3+C3) > 0.5 and A3/(A3+T3) < 0.5 regions, indicating a preference for T over A and G over C at the third codon position (Figure 6b).

The neutrality plot analysis, which assesses the relationship between GC content at the first two codon positions (GC12) and the third position (GC3), provided the most direct evidence of the dominant evolutionary force (Figure 6c). The GC12 values ranged from 0.40 to 0.56, while GC3 values ranged from 0.30 to 0.50, confirming the preference for A/T bases at the third position. The slopes of the regression lines fitted with GC12 and GC3 were consistently low, ranging from 0.13 to 0.30, and the $R^2$ values were also low, ranging from 0.028 to 0.144. These slopes deviate significantly from 1, indicating that neutral mutation pressure explains only a small proportion of the variance in CUB. This result strongly suggests that codon preference in the chloroplast genomes of these ferns is significantly influenced by natural selection. This finding is consistent with numerous studies across diverse lineages, from viruses to plants, that have used the same analytical methods to demonstrate the predominance of natural selection in shaping codon usage[32,37-40].

Although these analyses confirmed that natural selection shaped codon preferences in the chloroplast genomes of both Type A and Type C, they did not reveal clear distinctions in overall CUB patterns between the two types for all genes analyzed, including common genes such as *atpB*, *psaA*, *psbD*, *chlN*, *ndhA*, *ndhJ*, *ycf4*, and *psbA*. The distribution patterns in 2D scatter plots were similar, and neutrality plots did not show obvious differences in the regression results. This indicated that the significant separation of the two types observed in the codon UMAP landscape could not be reflected by overall differences in these general CUB metrics, but might reside in the special preferences of individual genes.

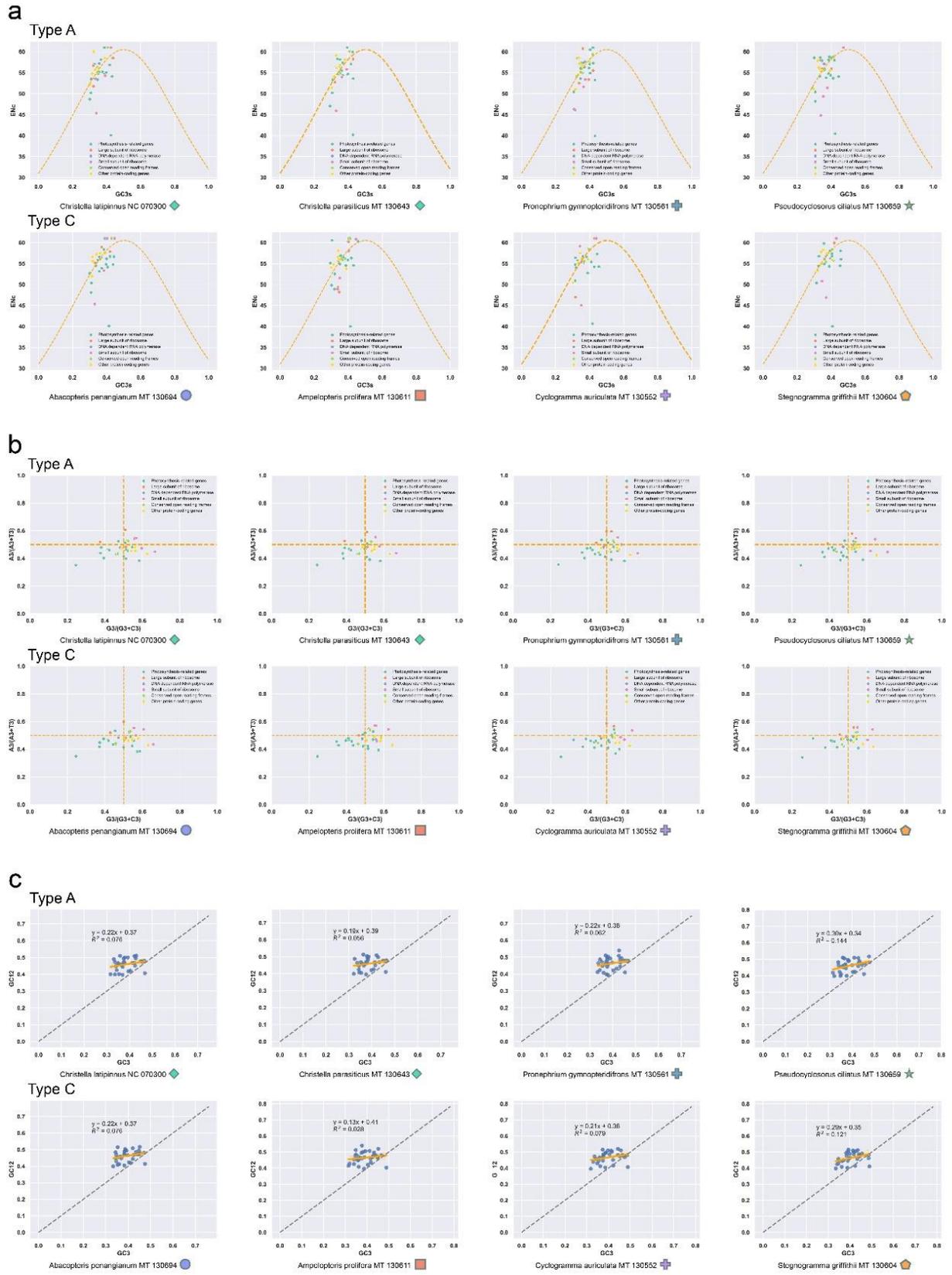

**Figure 6.** Analysis of Evolutionary Forces Shaping Codon Usage Bias in Thelypteridoideae. (a) Effective Number of Codons (ENC) plot. (b) PR2-plot. (c) Neutrality plot (GC12 vs. GC3).

## 6. The Convergent Signal is Driven by a Small Subset of Photosynthesis-Related Genes

Although the neutrality plot analysis confirmed that natural selection shaped CUB in both Type A and Type C, it did not reveal clear distinctions between the two types. To identify the specific genes responsible for the type separation observed in the UMAP analysis, we performed a gene-by-gene UMAP clustering using the Relative Synonymous Codon Usage (RSCU) values for each shared protein-coding gene. The results revealed that the UMAP clustering based on the RSCU values of three specific genes—*ndhJ*, *psaA*, and *psbD*—precisely recapitulated the global clustering pattern, individually segregating the species into the same distinct Type A and Type C clusters (Figure 7). In contrast, none of the other shared genes exhibited this clear separation pattern, strongly suggesting that the divergence in CUB between the two types is primarily driven by these specific genes. The functional roles of these genes, which are central to photosynthesis and stress response, are known to be under strong, dynamic environmental regulation, making them plausible targets for adaptive evolution[41-45].

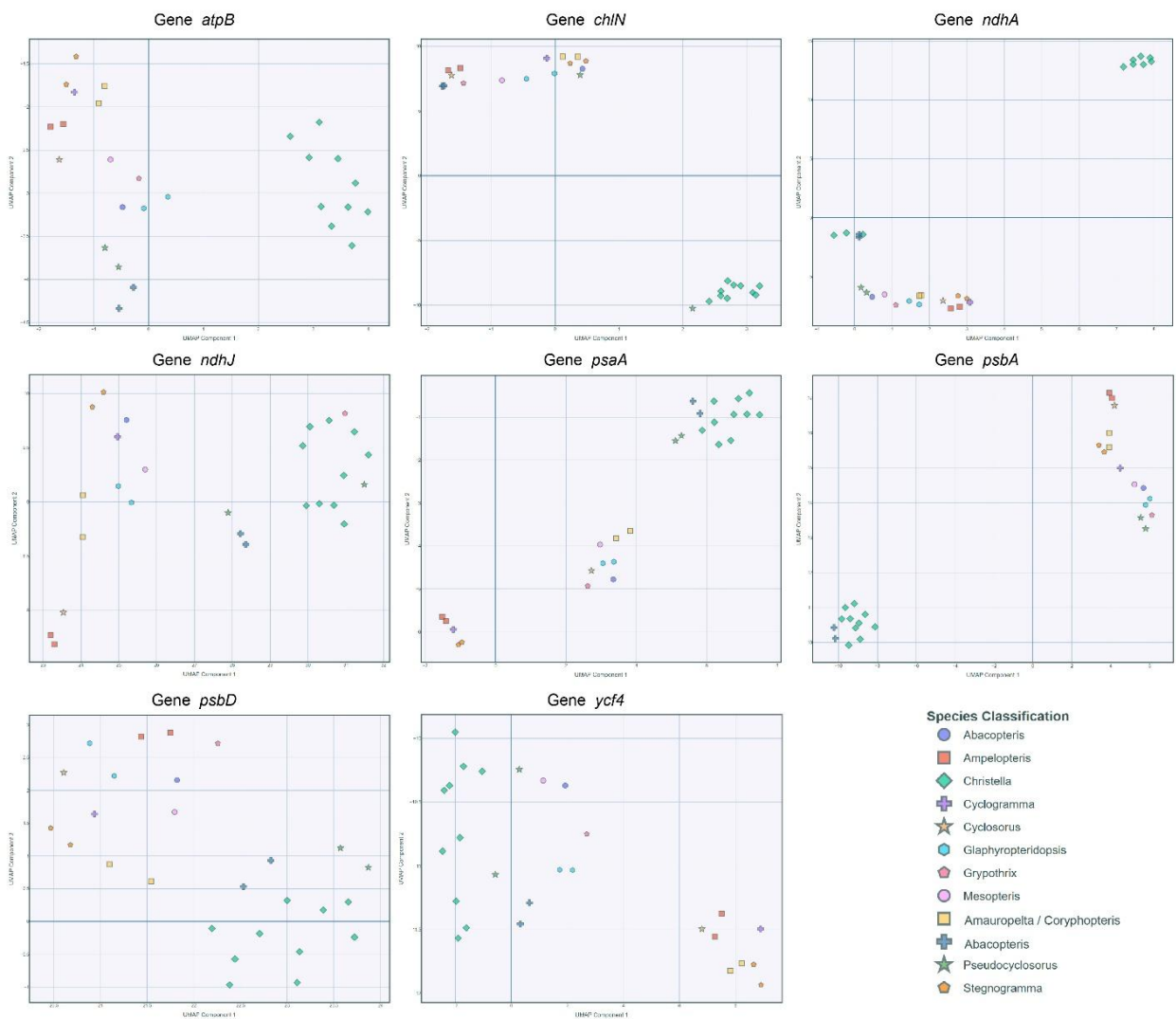

**Figure 7.** The gene-by-gene UMAP clustering using UMAP Visualization of Codon Usage Bias among Thelypteridaceae species of Type A and Type C.

To further validate whether these candidate genes possess sequence-level features that distinguish the codon usage patterns of Type A and Type C, we performed multiple sequence alignments for all shared genes using MAFFT. We then identified and compared the number of type-specific nucleotide sites and, more specifically, the number of type-specific sites located at the third codon position (i.e., codon bias sites) between the two types (Figure 8). The results showed that *ndhJ*, *psaA*, and *psbD* each contained a high number of type-specific sites (≥4) capable of distinguishing Type A from Type C, with *psaA* having the highest number (7). Moreover, these three genes also ranked among the highest in the number of codon bias sites, each having two or more such sites, with *psaA* again possessing the most. Crucially, the proportion of these codon bias sites relative to the total number of type-specific sites was exceptionally high in *psaA* (85.7%) and *psbD* (100.0%), far exceeding that of most other genes. The proportion in *ndhJ* was also substantial at 50%.

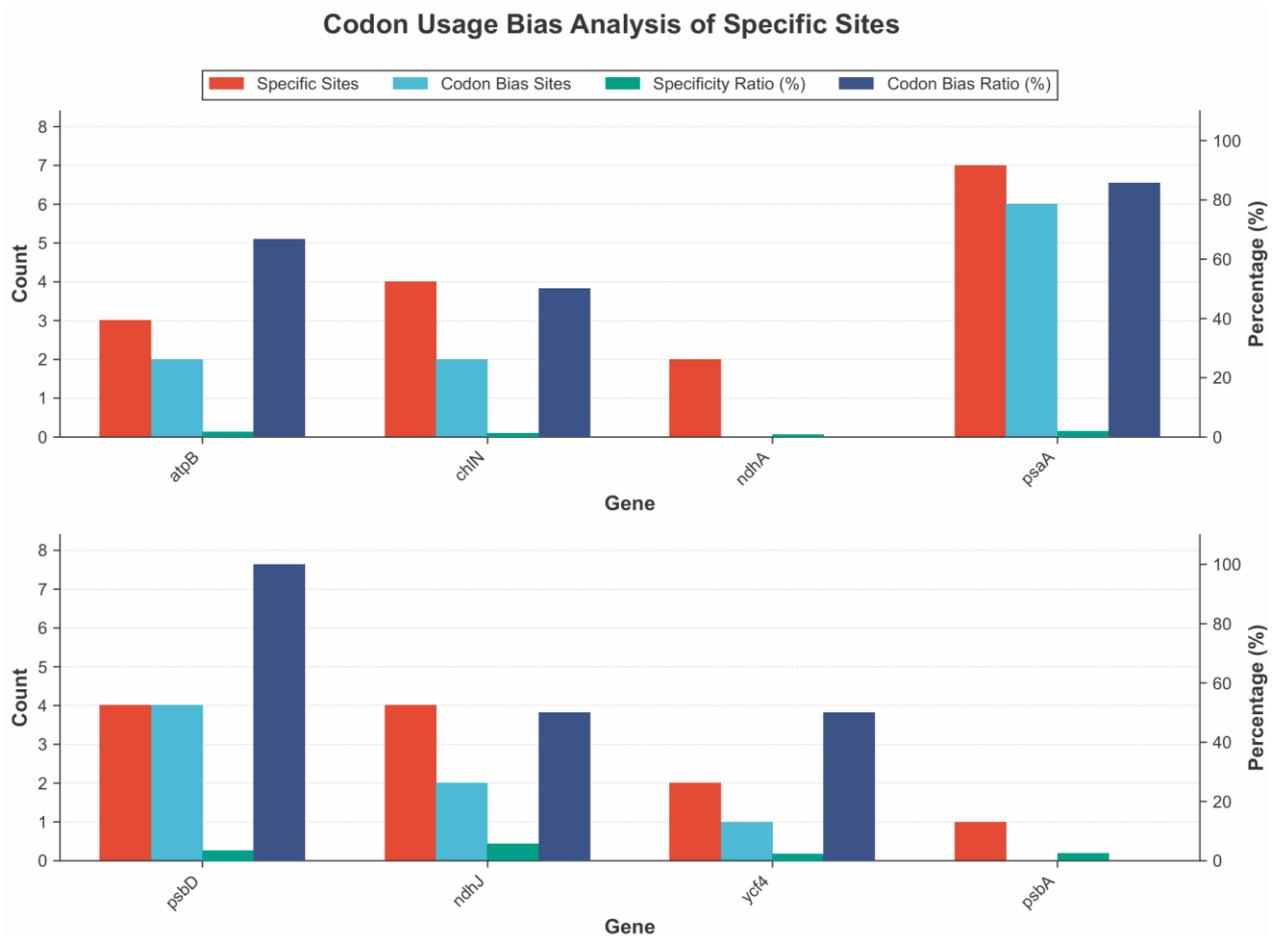

**Figure 8.** Sequence Alignment result of Type-Specific Codon Bias Sites for genes shared by Type A and Type C species. This figure provides a quantitative analysis of the diagnostic nucleotide sites that differentiate Type A and Type C across key chloroplast genes. For each gene, the chart displays the absolute count (left axis) of total type-specific nucleotide sites (red bars) and the subset of those sites located at the third codon position (codon bias sites; light blue bars). The right axis displays two ratios as percentages. The Specificity Ratio (green bars) indicates what proportion of a gene's total variable sites are type-specific. The dark blue bars (right axis) represent the ratio of codon bias sites relative to the total number of specific sites

Therefore, *ndhJ*, *psaA*, and *psbD* not only distinguish Type A and Type C at the statistical level of RSCU-based UMAP clustering but also at the sequence level, possessing a significantly higher number of diagnostic codon preference sites compared to other genes (Figure 8). This demonstrates that the distinctive codon usage patterns of these three genes are manifested not only in their overall RSCU statistical profiles but are also directly encoded at specific sites within their sequences. The abundance of type-specific sites, highly concentrated at the third codon position, provides direct molecular sequence evidence for the type separation observed in the UMAP analysis. This confirms that *ndhJ*, *psaA*, and *psbD* are key drivers of the codon usage bias differentiation between Type A and Type C.

In addition to these three genes, *atpB* and *chlN* also showed signals of differentiation (Figure 8). Both genes contained two codon bias sites, accounting for 66.7% and 50.0% of their respective type-specific sites. Although the RSCU-based UMAP clustering for these two genes did not perfectly resolve the Type A and Type C division, the separation was still discernible. The fact that these genes also exhibit diagnostic codon bias sites at the sequence level suggests that the selective pressures driving CUB divergence are not strictly confined to the three primary genes but may represent a more widespread evolutionary phenomenon. While the signal from *atpB* and *chlN* is weaker, preventing them from independently forming distinct clusters in the UMAP analysis, they serve as important supporting evidence for this evolutionary trend, thereby strengthening the reliability of our overall conclusion.

## Discussion

### An Adaptive Signal Overriding Phylogenetic History

This study provides strong evidence that codon usage bias (CUB) in specific chloroplast genes may have evolved convergently in concert with frond architecture in the fern family Thelypteridaceae. By integrating phylogenomics, advanced dimensionality reduction, and classical morphology, we have demonstrated that CUB can serve as a quantifiable molecular indicator of adaptive history, capable of overriding the background phylogenetic signal. Our analysis first established a robust chloroplast phylogeny that aligns with established classifications[6,28], confirming the major subfamily divisions. However, a Uniform Manifold Approximation and Projection (UMAP) analysis of codon usage parameters revealed a significant incongruence with this phylogeny, partitioning the species into distinct clusters that grouped phylogenetically distant genera. This incongruence was strongly explained by both its correlation with a key morphological trait (lamina base architecture[12]) and a distinct temporal origin for this trait revealed by our divergence time analysis. This finding provides new insight into our understanding of molecular convergence, extending it from the well-documented level of non-synonymous amino acid substitutions[1,2] to the more subtle, regulatory level of synonymous codon evolution. While previous studies have shown that CUB patterns are generally congruent with phylogeny[31,46], our results present a clear example where adaptive pressures have driven CUB to evolve in a manner that reflects functional convergence rather than shared ancestry.

### The Functional Basis of the Convergent CUB Signal

The functional significance of this molecular convergence is likely a consequence of the co-evolution of frond architecture and the regulation of the photosynthetic apparatus. The two distinct morphological syndromes identified—a non-truncate, tapering base (Type A) versus a truncate base (Type C)—represent different strategies for light capture and environmental interaction. Such morphological adaptations are known to be linked to ecological variables in ferns[7], We propose a functional hypothesis that these different architectures create distinct selective pressures on the photosynthetic machinery, requiring precise regulation of its expression. This hypothesis is strongly supported by our identification of *ndhJ*, *psaA*, and *psbD* as the primary drivers of the CUB divergence. These genes are not constitutively expressed housekeeping genes; they encode central components of the photosynthetic and stress-response pathways. The Ndh complex, which includes the NDH-J subunit, is critical for protecting against photo-oxidative stress under fluctuating environmental conditions[41], while the expression of *psaA* and *psbD* is known to be subject to strong, dynamic regulation in response to light and abiotic stress [42]. The convergent CUB patterns in these specific genes likely represent a form of molecular optimization to modulate their translational efficiency or accuracy[4,5], ensuring that the protein products are synthesized at the appropriate levels and times to meet the specific physiological demands of each morphological syndrome.

**Interpreting Convergence within a Complex Evolutionary History**

Our findings must be interpreted within the broader context of the complex evolutionary history of Thelypteridaceae, which is shaped by a combination of evolutionary processes. While our results provide powerful evidence for convergent evolution, the phylogenetic incongruence of the two *Christella parasiticus* accessions in our own dataset points to the influence of other significant forces. This observation is consistent with classical taxonomic studies that describe *C. parasiticus* as a highly variable species complex with multiple ploidy levels[30], and aligns with recent phylogenomic research demonstrating that intricate evolutionary patterns are common within the christelloid clade[10]. This underscores that the evolutionary history of Thelypteridaceae is not a simple bifurcating tree but a complex tapestry shaped by the interplay of vertical descent and, as we have shown, powerful convergent adaptation. Our divergence time estimation provides compelling corroboration for this adaptive narrative. The analysis places the radiation of the key Type A genera (*Abacopteris*, *Pseudocyclosorus*, and *Christella*) at approximately 19.43 Ma, during the early Neogene (Miocene). This period was characterized by significant global climatic shifts, including cooling, aridification, and the fragmentation of forest habitats[47,48]. Such environmental upheaval would have created a mosaic of new ecological niches, exerting strong selective pressures on understory plants and likely favoring novel adaptations for light capture and resource management. The emergence of the distinct tapering lamina base morphology of Type A within this specific geological timeframe strongly suggests it was an adaptive response to these changing conditions. The primary limitation of our study is its reliance on the chloroplast genome, which represents only the maternal lineage and cannot, by itself, fully resolve all sources of phylogenetic incongruence. Furthermore, while the correlation between CUB and lamina morphology is exceptionally strong, our proposed functional link remains a well-supported hypothesis that requires direct experimental validation. Lastly, our conclusion that *Christella* is monophyletic is valid for our sampled taxa but should be viewed with caution, as more comprehensive studies suggest the genus is paraphyletic when considered in its entirety 17[11,28],.

**Prospects for Future Investigation**

The results of this study open several promising avenues for future research. A critical next step will be to integrate nuclear genomic data with our chloroplast dataset. This will allow for a more complete understanding of the family's evolutionary history, helping to better resolve complex phylogenetic relationships and confirm the patterns of convergent evolution identified here. Furthermore, the functional hypothesis we propose can be tested experimentally. Synthetic biology approaches could be used to create versions of the *psaA* or *psbD* genes with the codon usage patterns characteristic of Type A and Type C, respectively. Expressing these constructs in a model system and measuring their translational efficiency and the host's photosynthetic performance under different light or stress conditions could provide direct evidence for the adaptive significance of the observed CUB patterns. Finally, the analytical framework developed here—using UMAP on codon usage parameters to detect non-phylogenetic patterns of adaptive convergence—represents a novel and effective method that could be applied to other plant and animal groups known for extensive morphological homoplasy to determine if the link between CUB and convergent adaptation is a general principle in molecular evolution.

**Conclusion**

This work demonstrates that the chloroplast genome contains a combination of evolutionary signals. We have shown that a subtle genomic feature, codon usage bias, can retain a clear and powerful signature of adaptive morphological convergence that is not apparent from phylogeny alone. By identifying a strong correlation between CUB patterns and lamina base architecture, pinpointing this radiation to a specific geological period, and identifying the specific photosynthesis-related genes driving this signal, our study provides a new framework for using such features to uncover the previously unobserved adaptive histories that shape organismal diversity.